\documentstyle[aps,epsf,rotate,subfigure]{revtex}
\font\mb=msbm10

\begin{document}
\draft
\title{Lyapunov Instability for a hard-disk fluid in equilibrium and
nonequilibrium thermostated by deterministic scattering} 
\author{C. Wagner}
\address{Center for Nonlinear Phenomena and Complex Systems,\\ 
Universit\'{e} Libre de Bruxelles, Campus Plaine Code Postal 231,\\
Boulevard du Triomphe, B-1050 Brussels, Belgium}
\date{\today}
\maketitle
\begin{abstract} 
We compute the full Lyapunov spectra for a hard-disk fluid under
temperature gradient and shear. The system is thermalized by
deterministic and time-reversible scattering at the boundary. This
thermostating mechanism allows for energy fluctuations around a mean
value which is reflected by only two vanishing Lyapunov exponents in
equilibrium and nonequilibrium. The Lyapunov exponents are calculated
with a recently developed formalism for systems with elastic hard
collisions. In a nonequilibrium steady state the average phase-space
volume is contracted onto a fractal attractor leading to a negative
sum of Lyapunov exponents. Since the system is driven inhomogeneously
we do not expect the conjugate pairing rule to hold which is confirmed
numerically.
\end{abstract}

\section{Introduction}
Transport of energy and momentum is a central problem in
nonequilibrium statistical mechanics, but so far most of our knowledge
is confined to the macroscopic level. There is still a long way to go
when it comes to understanding this phenomena in microscopic terms
although significant progress has been made during the last years,
with help from dynamical systems theory and computer simulations. In
general, external forces are needed to drive a system out of
equilibrium, but in order to prepare a nonequilibrium steady state the
redundant energy has to be removed thus preventing the system from
heating up indefinitely. One way out is the introduction of
thermostating mechanisms
\cite{Ev83,HLM82,EH83,Nose84a,Nose84b,Hoov85}. Both stochastic 
\cite{AT87,LeSp78,TCG82,GKI85,ChLe95,ChLe97,HoPo98,PoHo98} and
deterministic/time-reversible thermostats are in use, the latter
having been introduced to remain close to Hamiltonian dynamics.

Recently, an alternative thermostating mechanism
\cite{KRN99,RKN98,RKH99} acting via deterministic time-reversible
boundary-scattering has been applied on a hard disk fluid to model
heat and shear flow nonequilibrium steady states \cite{WKN99}. The
calculated transport coefficients have been found to be in agreement
with the theoretical values obtained from kinetic theory, but only for
special cases and in the thermodynamic limit the conjectured identity
between exponential phase-space contraction and entropy production
rate holds. In the present paper we investigate further the dynamical
properties of this system by computing the full Lyapunov spectra and
related quantities like the Kaplan-Yorke dimension or the
Kolmogorov-Sinai entropy. In Sec. \ref{sec1} we briefly recapitulate
the model and its thermostating mechanism and Sec. \ref{sec2} serves
to outline the method used for computing the Lyapunov exponents
\cite{DPH96,DePo97b}. The results are presented in Sec. \ref{sec3} and
conclusions are drawn in Sec. \ref{sec4}.

\section{Model}\label{sec1}
Consider a two-dimensional system of hard disks confined in a square
box of length $L$ with periodic boundary conditions along the x-axis,
i.e., the left and right sides at $x=\pm L/2$ are identified. The $N$
disks interact among themselves via elastic hard collisions, thus the
bulk dynamics is purely conservative. In the following and in all the
numerical computations we use reduced units by setting the particle
mass $m$, the disk diameter $\sigma$ and the Boltzmann constant $k_B$
equal to one. Now, denote with $p_x^i$, $p_y^i$ and with $p^f_x$,
$p^f_y$ the tangential and normal momentum of a disk before and after
a collision with the wall. Then the scattering prescription is given
as \cite{WKN99}
\begin{equation}\label{e1}
(p_x^f,p_y^f)=\left\{\begin{array}{ll}\mbox{\boldmath${\cal T}$
}^{-1}\circ{\cal M}\circ\mbox{\boldmath${\cal T}$ }  (p_x^i,p_y^i) ,&
\quad p^i_x\ge 0 \\ 
\Pi\circ\mbox{\boldmath${\cal T}$ }^{-1}\circ{\cal
M}^{-1}\circ\mbox{\boldmath${\cal T}$ }  (p^i_x,p^i_y) ,& \quad
p^i_x<0, 
\end{array}\right.
\end{equation} 
where $\mbox{\boldmath${\cal T}$ }:[0,\infty)\times[0,\infty)\to
[0,1]\times[0,1]$ is the invertible map
\begin{equation}\label{e3}
(\zeta,\xi) = \mbox{\boldmath${\cal T}$ }(p_x,p_y)
=\left(\mbox{erf}\left(|p_x|/\sqrt{2T}\right),\exp\left(-p_y^2/2T\right)\right)
\end{equation}
and ${\cal M}:[0,1]\times[0,1] \to[0,1]\times[0,1]$ is a
two-dimensional, invertible, phase-space conserving chaotic map to be
specified later. $\Pi(p_x,p_y)=(p_x,-p_y)$ only serves to produce the
right sign for the backward scattering. The parameter $T$ plays the
role of a temperature \cite{KRN99,RKN98,RKH99,WKN99}. Note that the
colliding disk retains its tangential direction and that the
scattering is reversible by construction.

So far, the model has only been defined in equilibrium. In order to
drive the system into a nonequilibrium steady state (NSS) the
collision rule only has to be modified appropriately, which will be
done in Sec. \ref{sec3.2} (see also \cite{WKN99}).

\section{Method}\label{sec2}
Like all hard disk systems our model is chaotic in the sense that two
nearby phase space trajectories diverge exponentially with time. This
is mainly due to the dispersing action of the hard disk collisions in
the bulk, but especially for small particle numbers we also expect a
contribution to this divergence due the chaotic nature of our
scattering mechanism. The average logarithmic divergence rate in phase
space are described by the so-called Lyapunov exponents
$\lambda_l$. Denote by ${\bf\Gamma}=\{{\bf q}_1,{\bf q}_2,...,{\bf
q}_N,{\bf p}_1,{\bf p}_2,...,{\bf p}_N\}$ the 4N dimensional phase
space vector for $N$ disks. Then, the time evolution
\begin{equation}\label{e4}
{\bf\Gamma}(0)=\Phi^t[{\bf\Gamma}(0)]
\end{equation}
of an initial state ${\bf\Gamma}(0)$ consists of a smooth streaming
which is interrupted by particle-particle and particle-wall
collisions. Next, consider a satellite trajectory ${\bf\Gamma}_s(t)$
initially displaced from the reference trajectory by an infinitesimal
vector $\delta{\bf\Gamma}(0)$. In a chaotic system
$|\delta{\bf\Gamma}(0)|$ is growing on average exponentially, thus
rendering the system unpredictable for long times. Then there exists a
complete set of linear-independent initial vectors
$\{\delta{\bf\Gamma}_l(0):l=1,\dots,4N\}$ and Lyapunov exponents
defined as \cite{Os68}
\begin{equation}\label{e5}
\lambda_l=\lim_{t\to\infty}\frac{1}{t}\ln\frac{|\delta{\bf\Gamma}_l(t)|}{|\delta{\bf\Gamma}_l(0)|}
.
\end{equation}
The $\lambda_l$, which we order according to
$\lambda_1\ge\lambda_2\ge\dots\ge\lambda_{4N}$, are independent of the
coordinate system and the metric. The whole set of Lyapunov exponents
is referred to as the Lyapunov spectrum.

In Hamiltonian systems the Lyapunov exponents appear in pairs summing
up to zero, $\lambda_i+\lambda_{4N-i+1}=0$ for $i=1,\dots,2N$, due to
the symplectic nature of the equations of motion. In a continuous
dynamical system one Lyapunov exponent associated with the direction
of the phase flow vanishes. Moreover, each conserved quantity leads to
an additional vanishing Lyapunov exponent. The symmetry found in
symplectic dynamical systems is lost when the system is driven to a
nonequilibrium stationary state. However, for homogeneous driving the
symmetry is replaced by the so-called {\it conjugate pairing rule}
\cite{ECM90} saying that after excluding the vanishing exponents
associated with the flow direction and the conservation of energy the
remaining pairs, i.e. $\{\lambda_1,\lambda_{4N}\}$,
$\{\lambda_2,\lambda_{4N-1}\}$, and so on, each sum up to the same
negative value $C$. For inhomogeneously driven systems such as ours or
the Chernov-Lebowitz shear flow model \cite{ChLe95,ChLe97,DePo97b},
however, the symmetry is lost and no pairing rules exist.

As can be seen from Eqs.(\ref{e1},\ref{e3}) the phase space volume is
in general changed during each disk-wall collision. In equilibrium
this averages up to zero, whereas in NSS the average phase space
contraction rate is negative and is given by the sum of all Lyapunov
exponents. Consequently, the phase volume shrinks continuously in NSS
and and the phase-space distribution collapses onto a multifractal
strange attractor. The fractal dimension of this strange attractor can
be estimated with the conjecture of Kaplan-Yorke \cite{KaYo79},
\begin{equation}\label{eky}
D_{KY}=n+\frac{\sum_{l=1}^n\lambda_l}{|\lambda_{n+1}|}
\end{equation}
where $n$ is the largest integer for which $\sum_{l=1}^n\lambda_l \ge
0$. $D_{KY}$ is the dimension of a phase space object which neither
shrinks nor grows and for which the natural measure is conserved by
the flow.

For the calculation of the full Lyapunov spectrum we use a method
worked out by Dellago {\it et al.}\cite{DPH96,DePo97b} which is
actually a generalization of the algorithm of Benettin {\it et
al.}\cite{BGGS80} for smooth dynamical systems. The latter follows the
time evolution of a reference trajectory and of a complement set of
tangent vectors by solving the original and the linearized equations
of motion, respectively. Periodic reorthonormalization prevents the
tangent vectors from collapsing all into the direction of fastest
growth. Averaging the logarithmic expansion and contraction rates of
the tangent vectors then yields the Lyapunov exponents. For a hard
disk system the free streaming is interrupted by impulsive collisions,
either with another particle or the boundary. This certainly affects
both the trajectory {\it and} the tangent space and has to be included
in the calculation. The free streaming and the particle-particle
collisions in the bulk have been treated in Section III-D of reference
\cite{DPH96}, where the same notation was used as in the present
work. We refer to Eqs. (39)-(42) and Eqs. (68)-(73) of this article
for explicit expressions of the particle-particle collision rules in
phase space and tangent space, respectively. It remains to consider
the particle-wall collisions and the following lines are formulated in
parallel to the respective treatment of Dellago and Posch for the
Chernov-Lebowitz model \cite{DePo97b}. In fact, the only difference
lies in the 'scattering matrix' and its derivatives.

If particle $k$ collides with the walls its position remains unchanged
whereas its momentum is changed according to the scattering rules
Eq. (\ref{e1}). The collision map ${\bf\Gamma}^f={\bf
M}({\bf\Gamma}^i)$ in phase space becomes
\begin{eqnarray}\label{e6}
{\bf q}^f_j&=&{\bf q}^i_j \quad \mbox{for}\quad j=1,\dots,N\\
{\bf p}^f_j&=&{\bf p}^i_j \quad \mbox{for}\quad j\ne k\\
{\bf p}^f_k&=&\mbox{\boldmath${\cal C}$ }^+({\bf p}^i_k),\quad
\mbox{for}\quad p^i_x \ge 0\\ 
{\bf p}^f_k&=&\mbox{\boldmath${\cal C}$ }^-({\bf p}^i_k),\quad
\mbox{for}\quad p^i_x<  0
\end{eqnarray}
using the abbreviations $\mbox{\boldmath${\cal C}^+$}=
\mbox{\boldmath${\cal T}$ }^{-1}\circ{\cal
M}\circ\mbox{\boldmath${\cal T}$ } $ for the forward scattering and
$\mbox{\boldmath${\cal C}^-$}= \Pi\circ\mbox{\boldmath${\cal T}$
}^{-1}\circ{\cal M}^{-1}\circ\mbox{\boldmath${\cal T}$ } $for the
backward scattering (Eq. (\ref{e1})).

In order to obtain the corresponding transformation for the tangent
space vector $\delta{\bf\Gamma}$ at a particle-wall collision we
assume that the collision takes place at phase point ${\bf\Gamma}$ at
time $\tau_c$. Then the satellite trajectory, displaced by the
infinitesimal vector $\delta{\bf\Gamma}$, collides at a different
phase point ${\bf\Gamma}+\delta{\bf\Gamma}_c$ at a different time
$\tau_c+\delta\tau_c$. A linear approximation in phase space {\it and}
time yields\cite{DPH96}
\begin{equation}\label{e7}
\delta{\bf\Gamma}^f=\frac{\partial {\bf M}}{\partial\bf\Gamma}\cdot
\delta{\bf\Gamma}^i + \left[\frac{\partial {\bf
M}}{\partial{\bf\Gamma}}\cdot{\bf F}({\bf\Gamma}^i)-{\bf F}({\bf
M}({\bf\Gamma}^i))\right]\delta\tau_c 
\end{equation}
where ${\bf F}$ is the right hand side of the equation of motion
during the free streaming
\cite{DPH96}, and $\partial {\bf M}/\partial{\bf\Gamma}$ is the matrix
of the derivatives of the full collision map with respect to the
phase-space coordinates. Obviously, the delay time $\delta\tau_c$ is a
function of the phase point ${\bf \Gamma}^i$ and of the tangent vector
$\delta{\bf \Gamma}^i$. For a disk-wall collision of the $k$th
particle the delay time $\delta\tau_c$ is given by
\begin{equation}\label{e8}
\delta\tau_c=-\frac{(\delta{\bf q}_k\cdot{\bf n})}{({\bf
p}_k/m\cdot{\bf n})}	. 
\end{equation}
Here, ${\bf n}$ is the normal vector of the wall pointing into the
simulation box. Since the scattering rules Eq. (\ref{e1}) for the
momentum components is independent of the position of the particle,
the matrix $\partial{\bf M}/ \partial{\bf\Gamma}$ has the form
\begin{equation}\label{e9}
\frac{\partial {\bf M}}{\partial{\bf\Gamma}}=\left(\begin{array}{cc}
{\bf 1}& {\bf 0}\\ {\bf 0}&\frac{\partial\mbox{\boldmath${\cal C}$
}^{\pm}({\bf p}^i)}{\partial({\bf p}^i)} 
\end{array}\right)
\end{equation}
where ${\bf 1}$ and ${\bf 0}$ are the $2N\times2N$ unit and zero
matrices, respectively. $\partial\mbox{\boldmath${\cal C}$
}^{\pm}({\bf p}^i)/\partial({\bf p}^i)$ is the matrix of the
derivatives of the outgoing momenta with respect to the incoming
momenta and only the components of the colliding particle $k$ are
different from zero. From Eq. (\ref{e7}) the following transformation
rules for the tangent vectors can be deduced:
\begin{eqnarray}\label{e10}
\delta{\bf q}^f_j&=&\delta{\bf q}^i_j \quad \mbox{for}\quad j\ne k\\
\delta{\bf p}^f_j&=&\delta{\bf p}^i_j \quad \mbox{for}\quad j\ne k\\
\delta{\bf q}^f_k&=&\delta{\bf q}^i_k-({\bf p}^f_k-{\bf p}^i_k)\delta\tau_c\\
\delta{\bf p}^f_k&=&\frac{\partial\mbox{\boldmath${\cal C}$
}^{\pm}({\bf p}^i_k)}{\partial({\bf p}^i_k)}\cdot\delta{\bf p}^i_k
. 
\end{eqnarray}
Omitting for notational convenience the index $k$ indicating the
colliding particle we obtain from Eq. (\ref{e1}) the following
expressions for the $2\times2$ matrix $(\partial\mbox{\boldmath${\cal
C}$ }^{\pm}({\bf p}^i_k)/\partial({\bf p}^i_k))_{\alpha\beta}=\partial
p_\alpha^f/\partial p_\beta^i$, $\alpha,\beta \in \{x,y\}$:
\begin{eqnarray}
\frac{\partial p_x^f}{\partial p_x^i}=\left({\bf D} {\cal
M}\right)_{11}\exp\left[((p_x^f)^2-(p_x^i)^2)/(2T)\right],&\quad&\frac{\partial
p_x^f}{\partial p_y^i}=-\left({\bf D} {\cal M}\right)_{12}\frac{\pi
p_y^i}{\sqrt{2T}}\exp\left[((p_x^f)^2-(p_y^i)^2)/(2T)\right]\label{e11}\\ 
\frac{\partial p_y^f}{\partial p_x^i}=-\left({\bf D} {\cal
M}\right)_{21}\frac{\sqrt{2T}}{\pi
p_y^f}\exp\left[((p_y^f)^2-(p_x^i)^2)/(2T)\right],&\quad&\frac{\partial
p_y^f}{\partial p_y^i}=\left({\bf D} {\cal
M}\right)_{22}\frac{p_y^i}{p_y^f}\exp\left[((p_y^f)^2-(p_y^i)^2)/(2T)\right]
,\label{e12}\\ 
\mbox{for}\quad p_x^i\ge 0 \nonumber	.
\end{eqnarray}
Here, ${\bf D}{\cal M}$ denotes the matrix of the derivatives of the
chaotic map ${\cal M}$. Eqs. (\ref{e11},\ref{e12}) are stated for
positive tangential velocities, for negative tangential velocities
${\cal M}$ has only to be replaced by ${\cal M}^{-1}$, see
Eq. (\ref{e1}).

Combining the free streaming with the transformation for the disk-disk
and the disk-wall collisions, one is now able to follow the exact time
evolution of the trajectory and of the tangent-space vector. 

\section{Results}\label{sec3}
Using the algorithm outlined in the previous section we are now able
to calculate the full Lyapunov spectrum for our hard disk model with
deterministic scattering at the boundary. As already mentioned we use
reduced units by setting the particle mass $m$, the disk diameter
$\sigma$ and the Boltzmann constant $k_B$ equal to unity. We define
the number density by ${\overline n}= N/L^2$. For simulation we use a
collision-to-collision approach and neighbor lists \cite{AT87}. For an
initial configuration the centers of the disks are positioned on a
triangular lattice and the momenta are chosen from a Gaussian with
zero mean. The total momentum is then set to zero and the momenta are
rescaled to obtain the total kinetic energy $E_{kin}=N(T_u+T_d)/2$,
$T_u$ and $T_d$ being the imposed 'parametrical' temperatures of the
upper and the lower wall, respectively.
\subsection{Equilibrium}
We now set both wall temperatures $T_u$, $T_d$ equal to one and
compute the full Lyapunov spectra for a four-particle system at number
density ${\overline n}=0.2$ using three different chaotic maps:
\begin{eqnarray}
{\cal
M}_B(\zeta,\xi)&=&\left(k\zeta,\xi/k\right)\hspace{2.25cm}\mbox{modulo
}1,\quad\mbox{(baker map)}\label{e13}\\ 
{\cal
M}_C(\zeta,\xi)&=&\left((k+1)\zeta+\xi,k\zeta+\xi\right)\quad\mbox{modulo
}1,\quad\mbox{(cat map)}\label{e14} 
\end{eqnarray}
and
\begin{equation}
{\cal M}_S:\left\{\begin{array}{l}\xi'=\xi-\frac{\displaystyle
k}{\displaystyle 2\pi}\sin(2\pi \zeta),\label{e15}\\ 
\zeta'=\zeta+\xi',\end{array}\right.\hspace{1.1cm}\mbox{modulo
}1,\quad\mbox{(standard map)} 
\end{equation}
with $0\le\zeta,\xi\le1$. $k\in2\hbox{\mb N}$ is a parameter
controlling the chaoticity of the map, i.e. the magnitude of the
Lyapunov exponents.

The resulting spectra are shown in Fig. \ref{fig1} where we have also
plotted the Lyapunov spectrum for elastic reflection as reference. To
emphasize the conjugate pairs, the Lyapunov exponents are ordered as
$\{\lambda_{2N-i+1},\lambda_{2N+i}\}$, with $i=1,\dots,2N$. Errors are
estimated as in \cite{DePo97b} from the convergence of the exponents as
a function of simulation time such that the time-dependent exponents
did not deviate more than $\pm\Delta\lambda$ from their mean values
during the second half of the simulation run. For high accuracy more
than $10^{7}$ disk-disk collisions and more than $5\cdot10^{6}$
disk-wall were simulated yielding errors less than $\pm 0.001$ for the
exponents and less than $\pm 0.002$ for the pair sums. In the case of
elastic reflection three Lyapunov exponents vanish. One exponent
vanishes due to the neutral expansion behavior in the direction of the
flow, a second due to the conservation of kinetic energy. The third
exponents is zero due to the translational invariance of the system in
the $x$-direction \cite{DePo97b}. The fourth vanishing exponent then
verifies the pairing rule. For a hard disk system thermostated by
deterministic scattering only two Lyapunov exponents vanish. The
kinetic energy is now allowed to fluctuate around a mean value, so
only the neutral expansion and the translational invariance remain. As
we expect the maximum Lyapunov exponent increases with increasing
chaoticity of the map, i.e. when going from a baker map ($k=2$) to a
cat map ($k=2$) to a standard map ($k=100$) (Results not plotted here
show a similar behavior when $k$ is increased for a given map.). The
pairing rule for these models is satisfied with an error of $\pm
0.002$ in the pair sums. At this point we add a remark which might
seem at first purely technical. For all simulations we used a
symmetrical configuration, i.e. Eq. (\ref{e1}) is used for the upper
wall whereas $\cal M$ and ${\cal M}^{-1}$ are interchanged in
Eq. (\ref{e1}) for the lower wall. Using the same scattering rules for
both walls results in an asymmetry and eventually in an asymmetric
Lyapunov spectrum even in equilibrium violating the pairing rule.

\subsection{NSS}\label{sec3.2}
We move on to the nonequilibrium stationary state and turn first to
the case of an imposed temperature gradient by the walls.  Since in
the thermodynamic limit the Lyapunov spectrum is mainly determined by
the bulk behavior we use in the following only a cat map with $k=2$ as
chaotic map $\cal M$. In order to determine the macroscopic state of
the system the velocity and density profiles of the bulk are measured
as well as the temperatures and velocities of the walls. Wall
velocities are defined as the mean tangential velocity of the incoming
and outgoing particles. Wall temperatures are defined as mean
temperature of the incoming and outgoing fluxes, ${\overline
T}=\left({\overline T}_i+{\overline T}_o\right)/2$, with ${\overline
T}_{i/o}=\left(\left<(v_x-\left<v_x\right>_x)^2\right>_x+\left[v_y\right]_y/\left[v_y^{-1}\right]_y
\right)/2$ where $\left<\,\right>_x$ and $[\, ]_y$ represent an
average over the density $\rho(v_x)$ and the flux $\Phi$ to and from
the wall, respectively (see also \cite{WKN99}).

\subsubsection{Heat flow}
Again, we first investigate a small system with four particles at
${\overline n}=0.2$ with high accuracy. In order to impose a
temperature difference on the system we only have to choose two
different parametrical temperatures $T_u$, $T_d$ (see Eq.(\ref{e3})
and \cite{WKN99}). Note that this also affects the derivatives of the
collision matrix (Eqs. (\ref{e11}, \ref{e12})).  Figure \ref{fig2}
shows the spectra for this system under a temperature gradient, the
numbers denoting the parametrical temperatures $T_{u/d}$ of the upper
and the lower wall. In NSS the sum of the Lyapunov exponents is
negative and exactly equal to the phase space contraction
rate. $T_u-T_d=1-3$ results in $\sum \lambda_l= -1.029$ and
$T_u-T_d=1-5$ in $\sum \lambda_l= -2.703$. We find again two vanishing
Lyapunov exponents but with increasing temperature difference all
nonzero exponents also increase in magnitude, the negative ones
certainly stronger to yield an overall negative sum. The pair sums are
also shown and, as we expect, the driving shifts the sums towards
negative values thus destroying the symmetry. The deviations from the
pairing rule are particularly strong for pairs with large $i$. Figure
{\ref{fig3} shows the results for a $36$-particle system under the
same setting. At least $2\cdot10^{6}$ disk-disk collisions and
$2\cdot10^{5}$ disk-wall collisions have been simulated in each
run. The spectra are plotted as connected lines only for graphical
reasons, it is understood that the exponents are defined for integer
$i$ only. The change in the Lyapunov spectrum under thermal driving
are similar to the four-particle system. Increasing the density from
${\overline n}=0.2$ to ${\overline n}=0.6$ results in a larger
magnitude of all nonzero exponents due to the higher collision
rate. The Kaplan-Yorke dimension$D_{KY}$ [Fig. \ref{fig4}] is
decreasing for increasing temperature gradient, with a larger
dimensionality loss $\Delta D_{KY}$ for higher densities than for
lower densities at given $\Delta \overline T$. As can immediately be
guessed from the positive branch of the spectra thermal driving also
results in an increasing Kolmogorov-Sinai entropy $h_{KS}$
[Fig. \ref{fig4}], defined as the sum over all positive exponents,
\begin{equation}\label{e16}
h_{KS}=\sum_{\{\lambda_l>0\}} \lambda_l\,,
\end{equation}
with increasing temperature gradient.  So, as we expect from
thermodynamics, thermal driving reduces the ordering of the system.
Higher collision rates at higher densities lead to more viscous
heating in the bulk and eventually to an increasing disorder
($h_{KS}$) of the system. The second, lower data point at $\Delta T=0$
shows $h_{KS}/N$ for elastic reflection as reference.

\subsubsection{Shear flow}
One way to model moving walls is to add some tangential momentum $d$
to $p_x$ before and after the collision of a particle with the
boundary (model I in \cite{WKN99}),
\begin{equation}\label{e17}
(p_x^f,p_y^f)=\left\{\begin{array}{ll}{\cal
S}_d\circ\mbox{\boldmath${\cal C}^+$}\circ{\cal S}_d\,(p_x^i,p_y^i) ,
&\quad p^i_x\ge -d \\ 
{\cal S}_d\circ\mbox{\boldmath${\cal C}^-$}\circ{\cal
S}_d\,(p^i_x,p^i_y) ,&\quad p^i_x<-d, 
\end{array}\right.
\end{equation} 
with
\begin{equation}\label{e18}
{\cal S}_d(p^i_x,p^i_y)=(p^i_x+d,p^i_y)\,.
\end{equation}
In order to impose shear the shift $d$ has only to be chosen with
different signs for the upper and the lower wall. Note that these
scattering rules are time reversible. Certainly, the drift also
affects the derivatives of the collision matrix,
Eqs. (\ref{e11},\ref{e12}), where $p_x^f$ goes to $p_x^f-d$ and
$p_x^i$ to $p_x^i+d$. Figure \ref{fig5} shows the full Lyapunov
spectra and the pair sums for a 36-particle system at ${\overline
n}=0.6$ under shear while keeping $T_u=T_d=1$ fixed. The negative
exponents increase in magnitude whereas the positive branch changes
very little. Before we take a closer look at the Kaplan-Yorke
dimension and the Kolmogorov-Sinai entropy let us investigate another
scattering rule (model III in \cite{WKN99}) which also models moving
walls:
\begin{equation}\label{e19}
(p_x',p_y')=\mbox{\boldmath${\cal T}$ }\!\!_*^{-1}\circ{\cal M}\circ\mbox{\boldmath${\cal T}$ }\!\!_*(p_x,p_y) 
\end{equation}
with
\begin{equation}\label{e20}
\mbox{\boldmath${\cal T}$ }\!\!_{*}(p_x,p_y)=\left(\frac{\mbox{erf}\left[(p_x-d)/\sqrt{2T}\right]+1}{2}
\, , \exp(-p_y^2/2T) \right)  
\end{equation}   
Model III is still deterministic but no longer time reversible and
only using this shear model a (numerical) equality between phase space
contraction rate and entropy production was found in
\cite{WKN99}. Note that $p_x^f$ changes now to $p_x^f-d$ and $p_x^i$
to $p_x^i-d$ in Eqs. (\ref{e11},\ref{e12}). Figure \ref{fig6} shows
the corresponding Lyapunov spectra for a 36-particle system at
${\overline n}=0.6$ under shear, but in contrast to model I the
Lyapunov exponents of both the positive and the negative branch now
increase in magnitude with increasing shear rate $\gamma$.  For
comparison the Lyapunov spectrum for the Chernov-Lebowitz model is
also plotted, with $E_{kin}/N$ and $\gamma$ equal to values obtained
with model III at $d=1.5$. $D_{KY}$ for both models is compared in
Fig. \ref{fig7a} and we see that the dimensionality loss with
increasing shear rate $\gamma$ is stronger for model I than for model
III.  The graph of $h_{KS}$, plotted for both models in
Fig. \ref{fig7b}, asks for more explanation. Firstly, the overall
behavior of $h_{KS}$ is increasing with larger $\gamma$, which seems
to be the opposite of the observation made for the Chernov-Lebowitz
shear model in \cite{DePo97b}. But there the total kinetic energy is
kept constant for all shear rates whereas here both models try to fix
the wall temperature. Increasing shear results in an increasing
viscous heat production in the bulk which is reflected by a larger
mean kinetic energy per particle [Fig. \ref{fig8a}]. Hence, the loss
in $h_{KS}$ due to the ordering introduced by the shear is more than
compensated by an increase of disorder due to a higher temperature in
the bulk. Secondly, the only minor changes in the positive branch of
the Lyapunov spectra for model I under shear yield an initially almost
constant or even decreasing Kolmogorov-Sinai entropy. This, and the
even more puzzling behavior of the wall temperature [Fig. \ref{fig8b}]
can be explained by the fact that model I does {\it not} produce a
Gaussian outgoing flux after the scattering. We found in \cite{WKN99}
that model I leads to an outgoing distribution with strong
discontinuities in NSS whereas model III yields proper outgoing
Gaussians. For comparison we have also computed $D_{KY}$ and $h_{KS}$
for the Chernov-Lebowitz model when it is approximately in the same
macroscopic state as model III, i.e. we set the Chernov-Lebowitz
system on the same kinetic energy shell and tried to find the
appropriate shear parameter which results in the same shear rate. The
Kaplan-Yorke dimension seems to be almost identical with that of model
I, Fig. \ref{fig7a}, and furthermore, the Kolmogorov-Sinai entropy now
also increases with larger shear rate, only differing by a constant
with the one obtained from model III, Fig. \ref{fig7b}. This offset,
depending on the special type of chaotic map chosen, originates from
the fluctuating character of the model and should vanish in the
thermodynamic limit.

\section{Conclusion}\label{sec4}
We have calculated the full Lyapunov spectrum for a hard-disk fluid in
equilibrium and nonequilibrium steady states thermostated by
deterministic scattering. Since the model allows for fluctuations
around a mean total energy only two vanishing Lyapunov exponents are
found in both equilibrium and nonequilibrium states. In nonequilibrium
the system is dissipative with a mean phase-space contraction rate
smaller than zero. The magnitude of the Lyapunov exponents increases
with increasing temperature gradient or shear rate, with a stronger
increase for the negative branch. Thus both heat and shear flow
situations result in a decreasing Kaplan-Yorke dimension and an
increasing Kolmogorov-Sinai entropy with stronger
nonequilibrium. Due to the inhomogeneous driving at the boundary the
pairing rule does not hold. We did not verify the relation between
Lyapunov exponents and transport coefficients (as e.g. in
\cite{ECM90}) since in absence of a pairing rule this would be
equivalent to checking the relation between phase space contraction
and entropy production rates, where the latter has been done in
\cite{WKN99}. The main difference of our shear flow model III and the
Chernov-Lebowitz shear flow model is the fact that in the latter the
total kinetic energy is fixed whereas our models tries to fix the wall
temperature. A direct comparison of both models reveals that they
yield identical Kaplan-Yorke dimensions and only differ by a constant
in the Kolmogorov-Sinai entropy, which vanishes in the thermodynamic
limit.
 
\section*{Acknowledgments}
Special thanks go to Ch. Dellago for providing and explaining the
original code used to compute the Lyapunov spectra and for helping to
interpret the results. Furthermore, the author likes to thank
O. Agullo for his technical support and R. Klages for pointing out
relevant literature and for giving valuables advice. This work is
supported, in part, by the Interuniversity Attraction Pole program of
the Belgian Federal Office of Scientific, Technical and Cultural
Affairs and by the Training and Mobility Program of the European
Commission.

\newpage
\begin{figure}
\epsfxsize=10cm
\centerline{\epsfbox{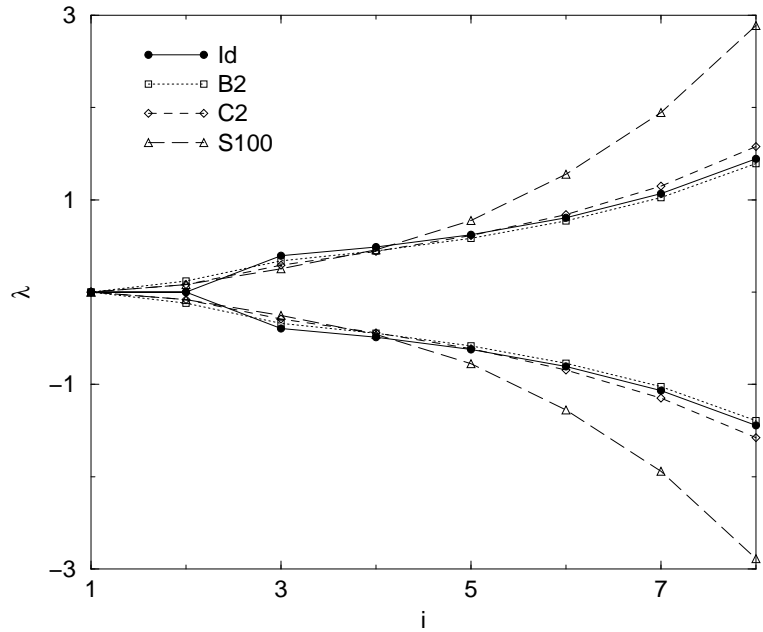}}
%\vspace*{0.3cm} 
\caption{Lyapunov spectra for a four particle system with density
${\overline n}=0.2$ in equilibrium. The solid line shows the spectrum
for elastic reflection, i.e. the identity map is used instead of a
chaotic one. The other spectra are obtained by using a baker map with
$k=2$, a cat map with $k=2$ and a standard map with $k=100$, see
Eqs.(\ref{e13})-(\ref{e15}).} 
\label{fig1}
\end{figure}

\begin{figure}
\epsfxsize=10cm
\centerline{\epsfbox{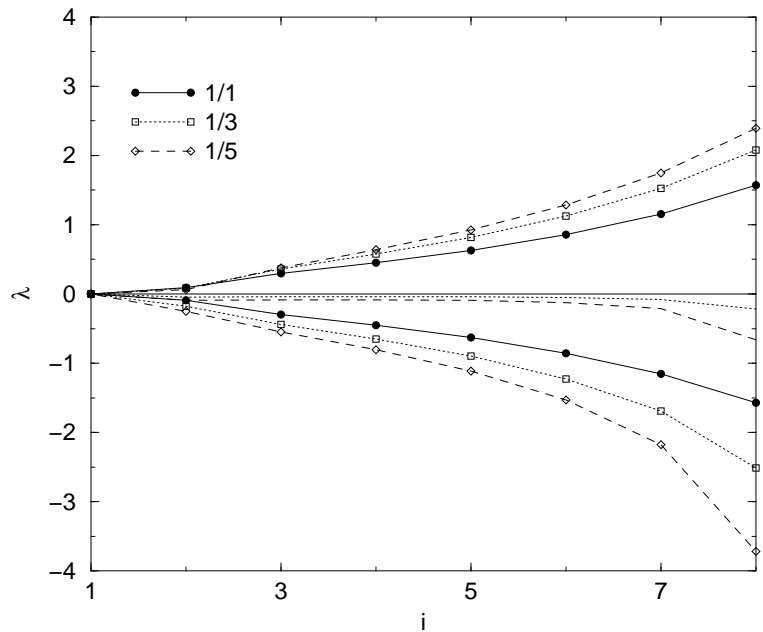}}
%\vspace*{0.3cm} 
\caption{Lyapunov spectra for a four particle system at density
${\overline n}=0.2$ in NSS. The imposed temperatures for the upper and
the lower wall are indicated by the numbers. The respective pair sums
(scaled by a factor $1/2$ for graphical reasons) are also plotted near
the middle line.} 
\label{fig2}
\end{figure}

\begin{figure}
\begin{center}
\epsfxsize=12cm
\epsfysize=9cm
\subfigure[]{\epsfbox{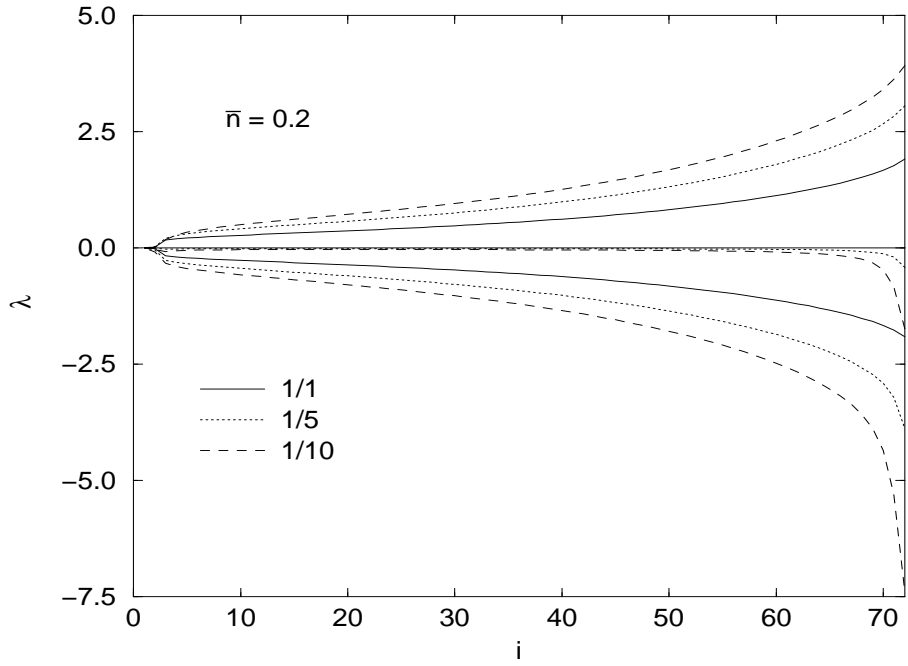}}\\
\epsfxsize=12cm
\epsfysize=9cm
\subfigure[]{\epsfbox{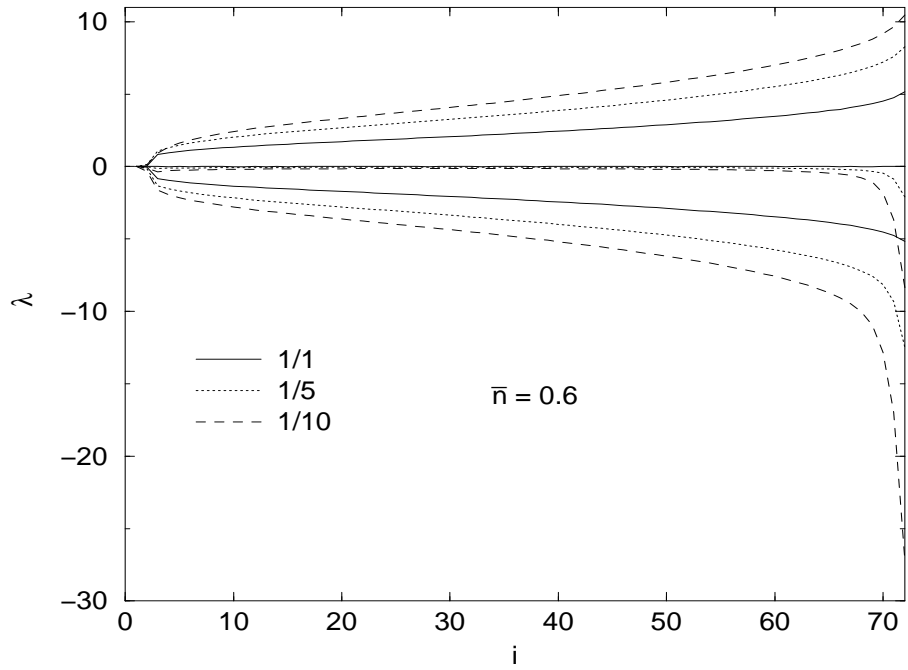}}
\end{center}
\caption{Full Lyapunov spectra for a $36$-particle system under an
imposed temperature gradient at number densities ${\overline n}=0.2$
and ${\overline n}=0.6$. The numbers indicate the parametrical
temperatures $T_u$, $T_d$. The respective pair sums (scaled by a
factor $1/2$ for graphical reasons) are also plotted near the middle
line.} 
\label{fig3}
\end{figure}

\begin{figure}
\begin{center}
\epsfxsize=12cm
\epsfysize=9cm
\subfigure[]{\epsfbox{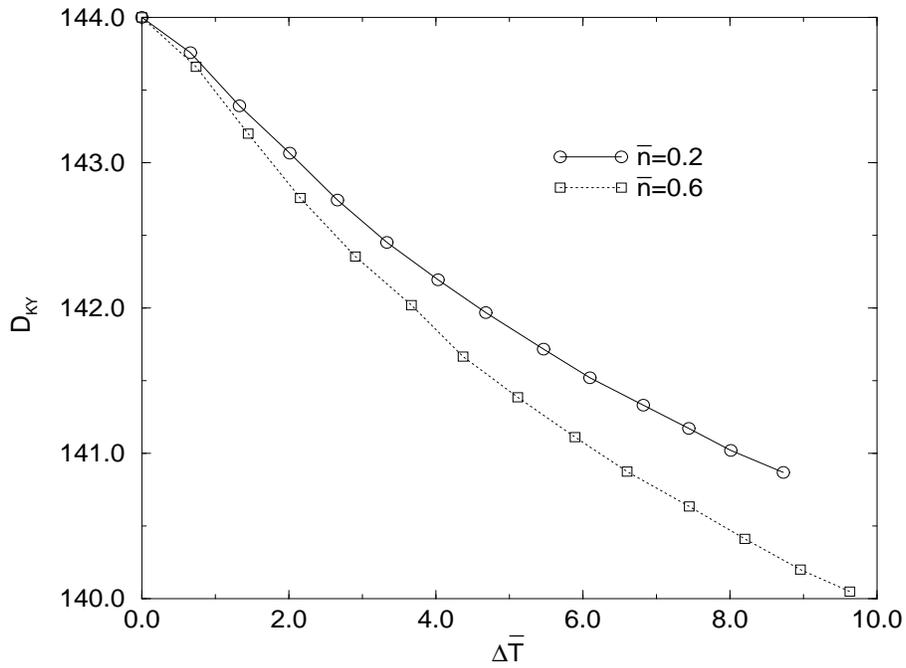}}\\
\epsfxsize=12cm
\epsfysize=9cm
\subfigure[]{\epsfbox{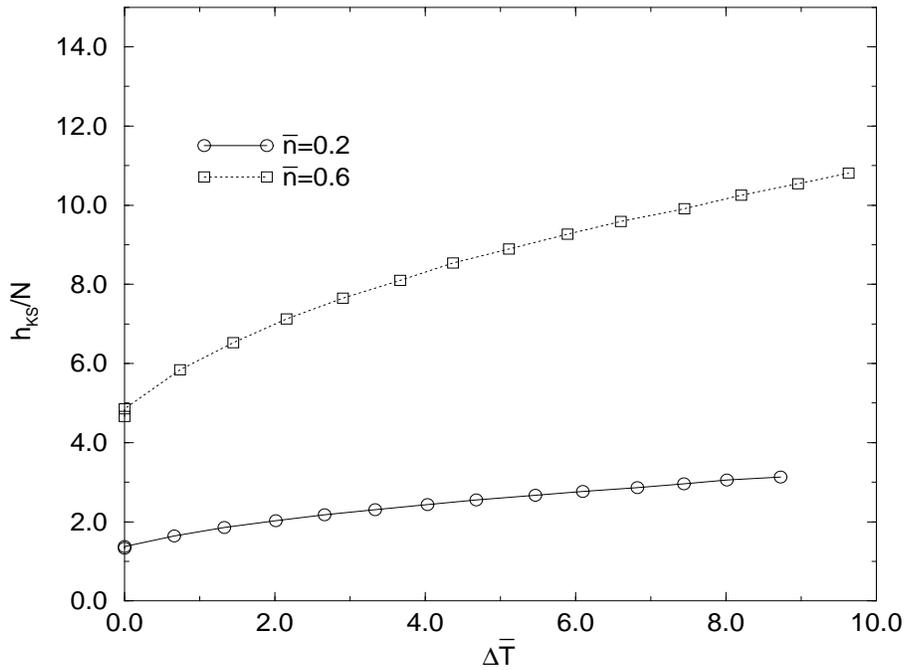}}
\end{center}
\caption{The Kaplan-Yorke dimension $D_{KY}$ and the Kolmogorov-Sinai
entropy $h_{KS}$ per particle for a 36-particle system under a
temperature gradient at number densities ${\overline n}=0.2$ and
${\overline n}=0.6$. $\Delta {\overline T}$ denotes the measured
temperature difference between the upper and the lower wall. The
second, lower data point at $\Delta {\overline T}=0$ gives $h_{KS}/N$
for elastic reflection as reference.}  
\label{fig4}
\end{figure}

\begin{figure}
\epsfxsize=12cm
\epsfysize=9cm
\centerline{\epsfbox{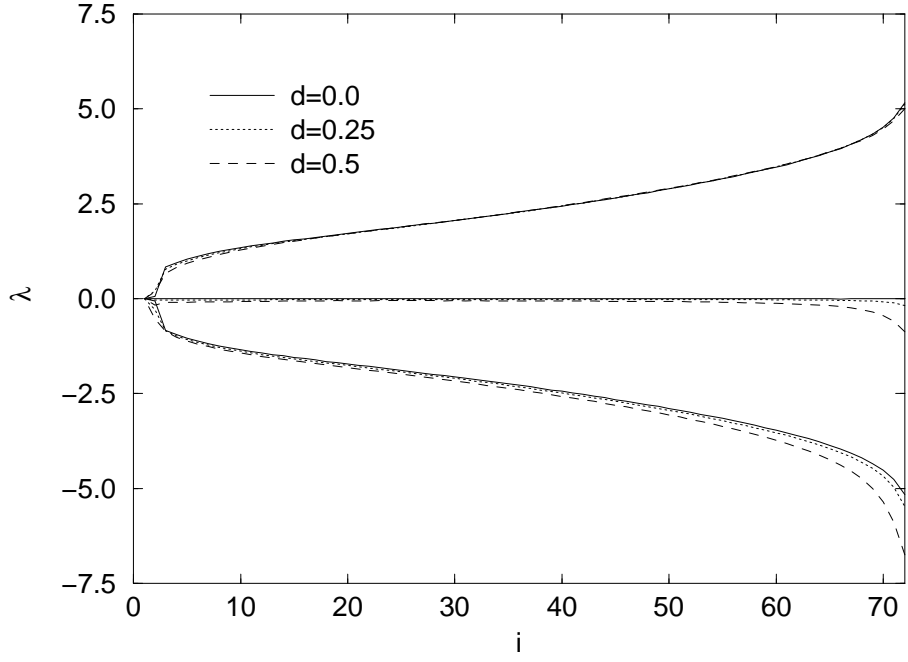}}
%\vspace*{0.3cm} 
\caption{Full Lyapunov spectra for a $36$-particle system under shear
at number density ${\overline n}=0.6$, model I} 
\label{fig5}
\end{figure}

\begin{figure}
\epsfxsize=12cm
\epsfysize=9cm
\centerline{\epsfbox{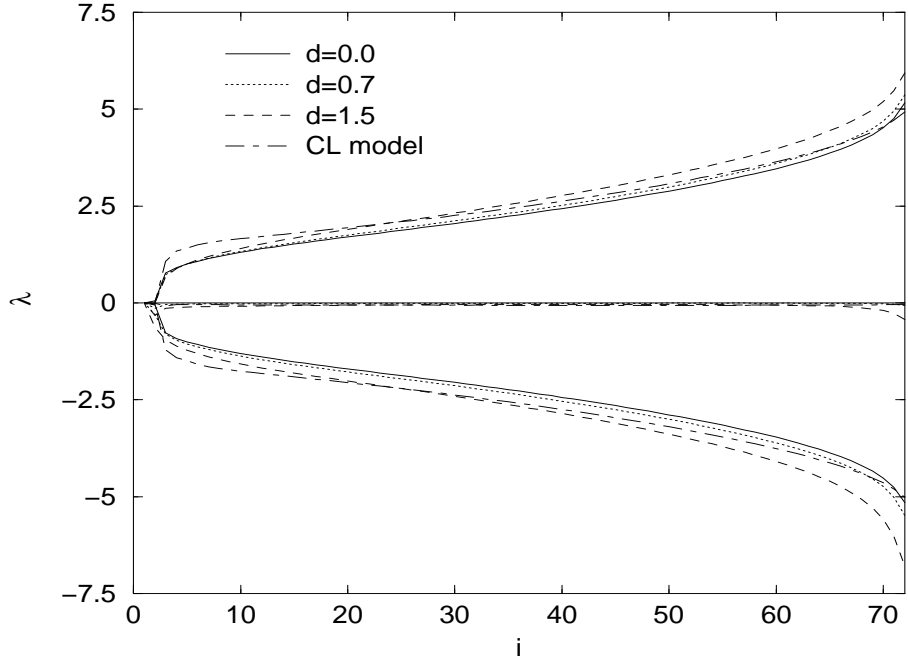}}
%\vspace*{0.3cm} 
\caption{Full Lyapunov spectra for a $36$-particle system under shear
at number density ${\overline n}=0.6$, model III. The spectrum for the
Chernov-Lebowitz model results from a simulation with the same mean
kinetic energy per particle and approximately the same shear rate as
model III with $T_u=T_d=1.0$ and $d=1.5$. } 
\label{fig6}
\end{figure}

\begin{figure}
\begin{center}
\epsfxsize=12cm
\epsfysize=9cm
\subfigure[\label{fig7a}]{\epsfbox{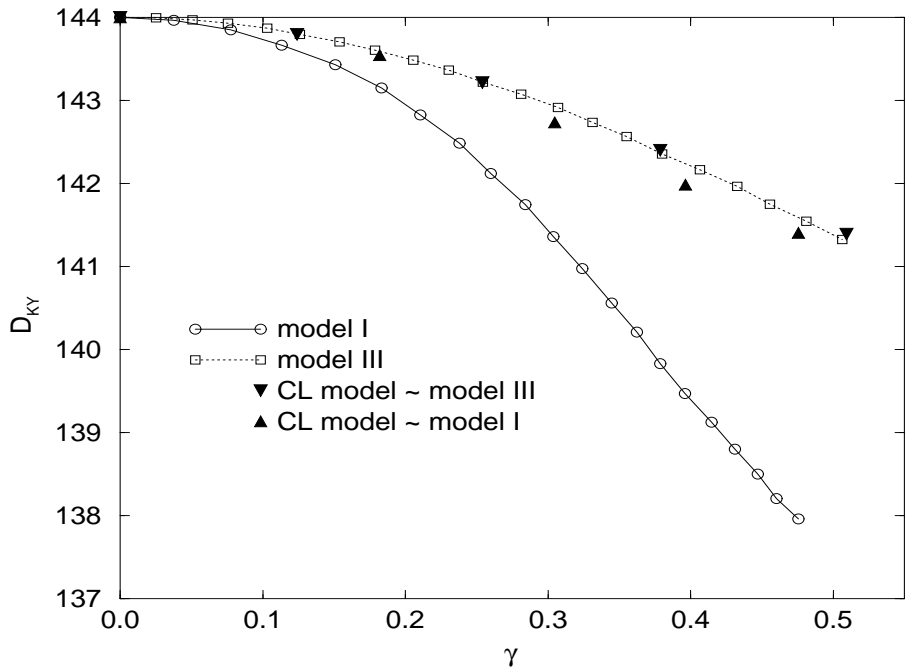}}\\
\epsfxsize=12cm
\epsfysize=9cm
\subfigure[\label{fig7b}]{\epsfbox{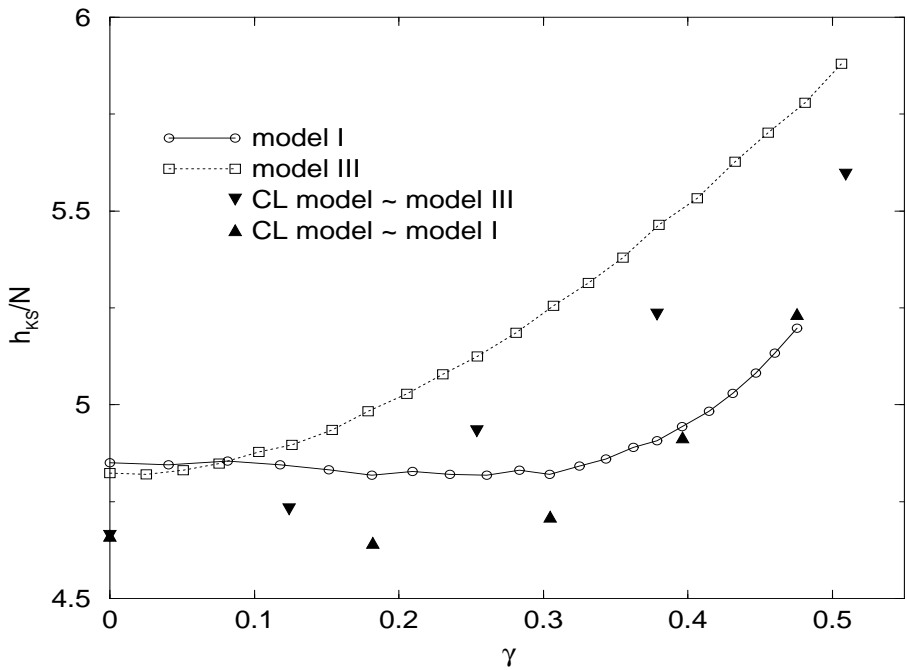}}
\end{center}
\caption{The Kaplan-Yorke dimension $D_{KY}$ and the Kolmogorov-Sinai
entropy $h_{KS}$ per particle for a 36-particle system under shear at
number density ${\overline n}=0.6$  as a function of the shear rate
$\gamma$ (model I, model III and Chernov-Lebowitz model).}  

\end{figure}

\begin{figure}
\begin{center}
\epsfxsize=12cm
\epsfysize=9cm
\subfigure[\label{fig8a}]{\epsfbox{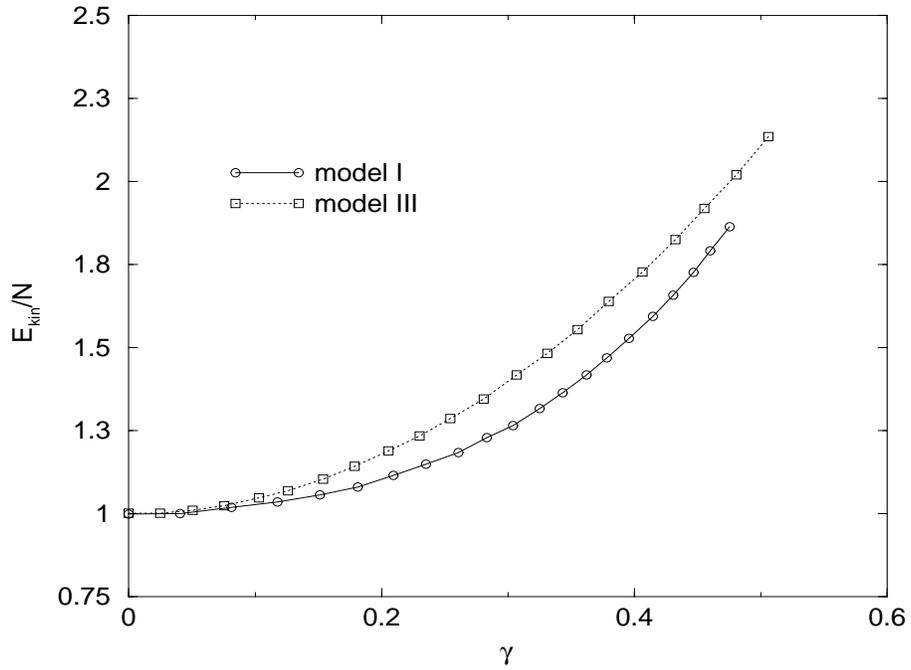}}\\
\epsfxsize=12cm
\epsfysize=9cm
\subfigure[\label{fig8b}]{\epsfbox{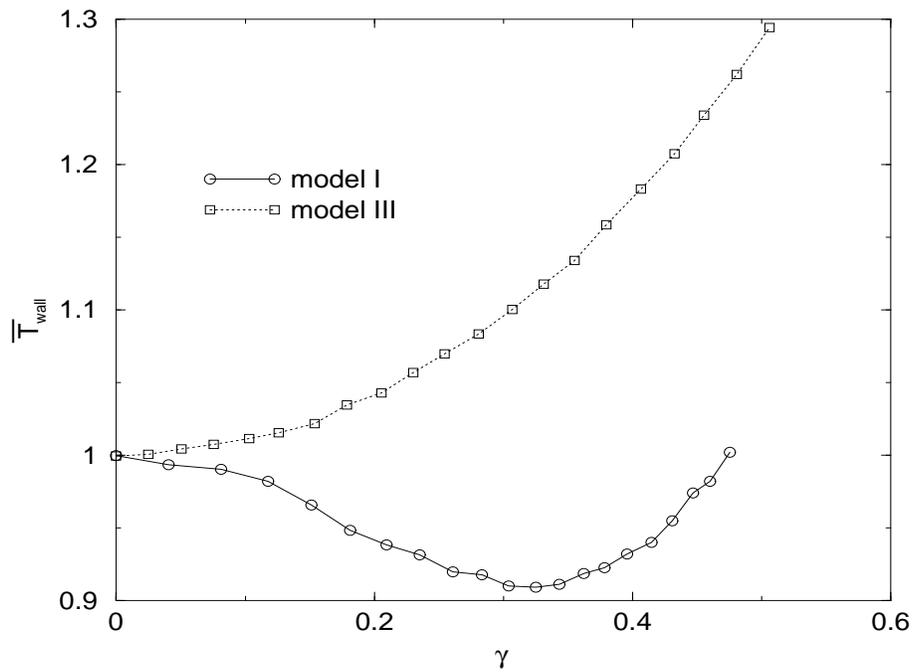}}
\end{center}
\caption{Kinetic energy per particle and measured wall temperature for
a 36-particle system under shear at number density ${\overline n}=0.6$
as a function of the shear rate  $\gamma$ (model I and model III).}  
\end{figure}

\end{document}